\documentclass{moriond}

\def\Journal#1#2#3#4{{#1} {\bf #2}, #3 (#4)}

\def\NIM{\em Nucl. Instrum. Methods}

\def\PLB{{\em Phys. Lett.}  B}

\def\PRD{{\em Phys. Rev.} D}

\def\be{\begin{equation}}
\def\ee{\end{equation}}
\def\bea{\begin{eqnarray}}
\def\eea{\end{eqnarray}}

\newcommand{\Photo}{}

\begin{document}
\vspace*{4cm}
\title{Results from the NA62 2014 Commissioning Run}

\author{ DARIO SOLDI 
\\
On Behalf NA62 Collaboration \footnote{
G.~Aglieri Rinella, R.~Aliberti, F.~Ambrosino, B.~Angelucci, A.~Antonelli, G.~Anzivino, 
R.~Arcidiacono, I.~Azhinenko, 
S.~Balev, J.~Bendotti
% ...
}
}

\address{Universita' degli Studi di Torino, Dipartimento di Fisica,\\
Via Pietro Giuria 1, Torino}

\maketitle\abstracts{
The main purpose of the NA62 experiment is to measure the branching ratio of the (ultra) rare decay $K^+ \rightarrow \pi^+ \nu \bar{\nu}$ with the precision of 10\% collecting $\sim$ 100 events with the Standard Model branching fraction in 3 years of data taking. 
The commissioning of the experiment after the 2014 pilot run and the prospects for the 2015 run are presented.  
}

\section{Introduction}
The NA62 experiment is located at the CERN Super Proton Synchrotron accelerator. Its main goal is to measure the branching ratio (BR) of the ultra rare decay \cite{Anelli} $K^+ \rightarrow \pi^+ \nu \bar{\nu}$, with the precision of $\sim$ 10 \%.
The Standard Model (SM) prediction is very accurate, taking in account next-to-leading order (NLO) QCD corrections to the top quark contributions, NNLO QCD corrections to the charm contributions, NLO electroweak corrections to both top and charm contributions and extensive calculations of isospin breaking and non-perturbative effects. The best prediction of the branching ratio \cite{Brod} is $BR(K^+ \rightarrow \pi^+ \nu \bar{\nu}) = (7.81^{+0.8}_{-0.71} \pm 0.29) \times 10^{-11}$  while the experimental measurement, combining data from E787 and E949 experiments \cite{E949} at Brookhaven AGS is $(17.3^{+11.5}_{-10.5}) \times 10^{-11}$ .  
\\
During 2014 the NA62 detector was deployed and commissioned, enabling to accumulate a valuable sample of kaon decays to verify the expected physics sensitivity.

\subsection{Experimental strategy}\label{subsec:exp}
\begin{figure}[h!]
\centering
\includegraphics[width=0.4\linewidth]{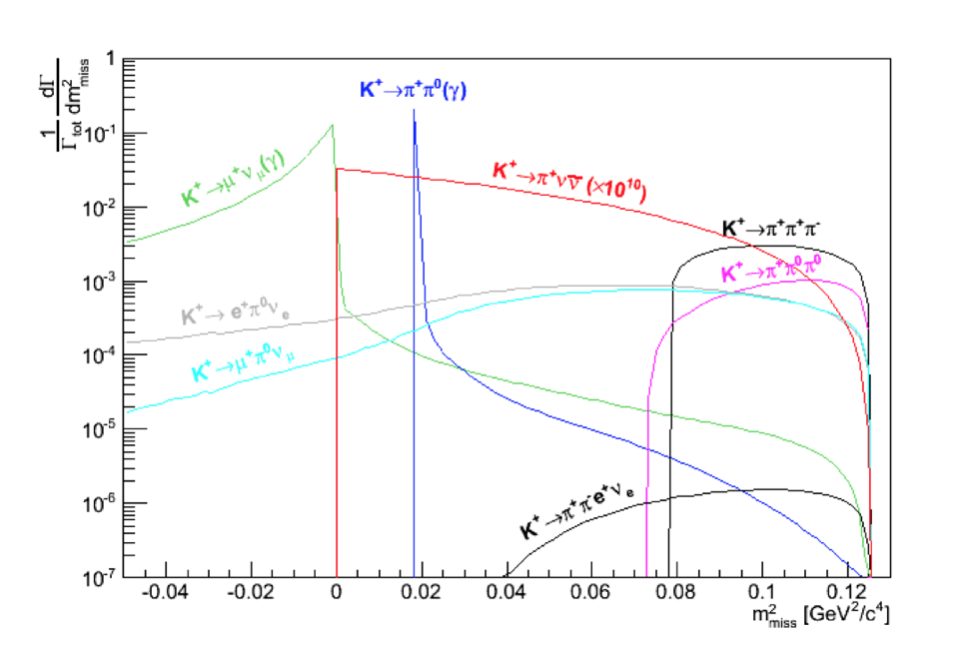}
\caption[]{squared missing mass in the hypothesis of kaon to pion decay}
\label{mm2_expected}
\end{figure}

A large number of about $10^{13}$ kaon decays is needed to reach the expected sensitivity. 
The experimental principle is the decay-in-flight technique, and the variable used to distinguish the signal from the background is the squared missing mass $m^2_{miss} = (P_{K^+} - P_{\pi^+})^2$ where $P$ are the four momenta of the incoming kaon and outgoing pion. Cuts on the missing mass value allow the rejection of more than 90\% of the backgrounds.
Figure \ref{mm2_expected} shows the $m^2_{miss}$ distribution assuming that the charged particle is a pion with momentum lower than 35 GeV. One defines two regions where the signal is not dominated by background, Region I above the $K_{\mu2}$ but below the $K_{\pi^+\pi^0}$ contribution and Region II above the $K_{\pi^+\pi^0}$ and below the $K_{3\pi}$ contributions.
The final rejection relies on a redundant veto system and particle identification.

\section{Detector and Beam during the 2014 commissioning run}
\subsection{Beam}\label{subsec:beam}
Kaons are produced by the 400 GeV protons from the SPS impinging on a beryllium target. Secondary particles of 75 GeV/$c$ momentum are selected in a non separated beam composed of 71\% of pions, 23\% of protons and 6\% of kaons, with a total rate of $\sim$ 750 MHz. Only about 10\% of the kaon produced decay in the fiducial volume, allowing the detector to collect $\sim 4.5~ 10^{12}$ decays per year.
\\
During the 2014 run the beam optics has been fully commissioned, and the experiment was operating at 5\% to 20\% of the nominal intensity.
   
\subsection{Kaon Tagging}\label{subsec:ktag}
Kaon tagging is performed by the CEDAR/KTAG detector. CEDAR \cite{CEDAR} is a differential Cerenkov counter built for measuring the composition of SPS secondary beams. It has been upgraded to meet the 50 MHz kaon rate requirement: a time resolution better than 100 ps and a tagging efficiency above 95\%.
\\
During the 2014 run a preliminary evaluation of the single hit time resolution of $\sim$ 280 ps was was obtained, while the tagging efficiency was above 90\%.

\subsection{Spectrometers}\label{subsec:spectro}
Both the beam particle and the secondary charged particle must be accurately measured. 
The beam spectrometer, called GigaTracker \cite{GTK}, consists of three stations of hybrid silicon pixel detectors installed within an achromatic system of four dipole magnets. The constraint of very low material budget has led to the choice of a sensor thickness of 200 $\mu$m, read by 10 ASIC 100 $\mu$m thick chips bump bonded to the sensor, with a total of 18000 pixels for each station. The time resolution per track should be of the order of 200 ps, while the momentum resolution $\sigma (P)/P$  should be $\sim$0.2\%. 
\\
During the commissioning run, the three stations have been installed, but only partially read out (10\% of the chips). The time resolution measurement is in progress.
\\
The Straw Tracker detector \cite{STRAW} (STRAW) measures the direction and momentum of charged particles from the kaon decay.
It is composed of 4 chambers with x,y,u,v views, located in vacuum, before and after a dipole magnet which produces the vertical B-field of 0.36 T. 
\\
The detector was fully operational.
\\
Both spectrometers were read out  triggerless and the matching with the other detectors was performed offline.

\subsection{Photon Vetoes}\label{subsec:PV}
The photon veto system was designed to suppress the background coming from $K^+ \rightarrow \pi^+\pi^0 ~(\pi^0 \rightarrow \gamma\gamma)$ and radiative decays by identifying photons with an inefficiency less than $10^{-8}$. The system provides a hermetic coverage for photon angles up to 50 mrad.
\\
The Large Angle Vetoes (LAV) cover the region between 8.5 and 50 mrad. The LAV consists of 12 stations, made of rings of lead glass blocks recovered from the OPAL electromagnetic calorimeter \cite{opal}.
\\
All stations have been installed and read out, and the time resolution of tagged photons measured is better than 1 ns.
\\
The other photon veto used in 2014 is the Liquid Krypton calorimeter (LKr), inherited from the NA48 experiment and equipped with a new readout. It covers photon angle from 1.5 to 8.5 mrad and should guarantee an inefficiency lower than $10^{-5}$ for photons with energy $> 35$ GeV.
\\During the 2014 data taking, the new read out was deeply tested and a calibration of the detector was performed.

\subsection{RICH counter}\label{subsec:RICH}
The Ring-Imaging Cherenkov Counter \cite{RICH} (RICH) is located between the last STRAW chamber and the LKr.   
It is a 17 m long vessel filled with Neon at atmospheric pressure, and consists of a mosaic of spherical mirrors which reflect the Cherenkov light on to the 2000 photomultiplier array.
The RICH provides $ \pi - \mu$ separation for charged particles with momentum between 15 and 35 GeV/$c$ leading to an additional muon suppression factor of $10^2$. It also measure the pion crossing time with about 100 ps precision.
\\
With the data taken during the 2014 run it is possible to test the ring reconstruction and the matching with the montecarlo expectation.  
\subsection{Muon Vetoes}\label{subsec:MV}
The muon veto system (MUV) consists of three different components (MUV1, MUV2 and MUV3) placed after the LKr calorimeter. The MUV3 component is a real muon veto, consisting of an array of 144 scintillator tiles. The two others, MUV2 and MUV3, are hadronic calorimeters (Fe-scintillators), used to measure the deposited energies and shower shapes of incident particles and redundantly distinguish muons from pions.
\\
During the commissioning run MUV2 and MUV3 were installed and read out. In particular MUV3 has been used for muon rejection at the hardware trigger level. The efficiency of MUV2 has been measured to be $\sim 99.9 \%$ with a preliminary time resolution of 1.4 ns.
\section{Trigger and Data Acquisition}
All the acquisition boards were installed for the 2014 run. The electronic boards were deeply tested under the available beam condition (5 to 20\% of the nominal intensity).
\\
Two level-0 trigger processors have been developed, one completely based on a FPGA commercial board, while the other \cite{L0TP} makes use of an auxiliary PC to build the trigger logic. 
\\For the 2014 run the first option has been used.

\section{First look at the data}
During the last days of the 2014 run, data in stable conditions at about 5\% of the nominal intensity have been recorded as selected by a Level-0 trigger logic based on fast multiplicity signals from the Charged Hodoscope (CHOD), hadronic energy measured in MUV2 and requiring little or no electromagnetic energy deposited in the scintillating fibre hodoscope embedded in the LKr calorimeter. 
The analysis of the data sample is in progress \cite{SPSC}. 
\begin{figure}[h]
\begin{minipage}{0.50\linewidth}
\centerline{\includegraphics[width=0.7\linewidth]{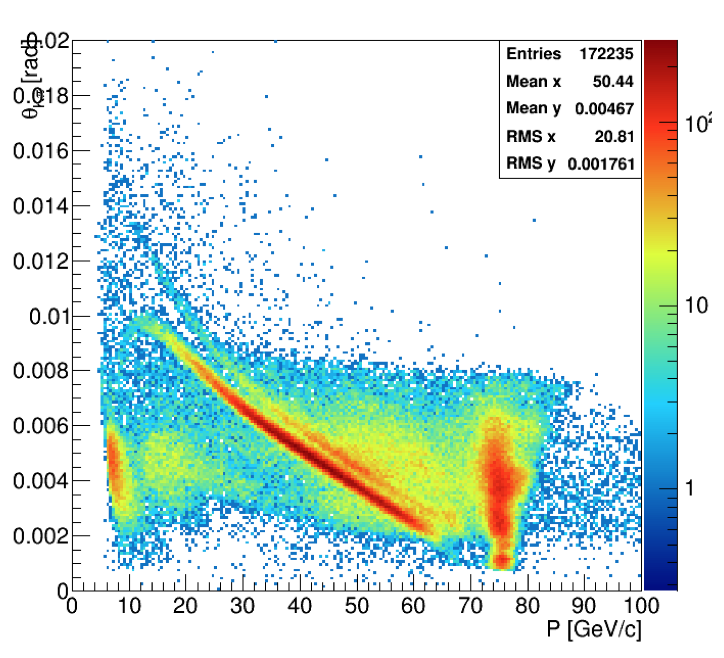}}
\end{minipage}
\hfill
\begin{minipage}{0.50\linewidth}
\centerline{\includegraphics[width=0.7\linewidth]{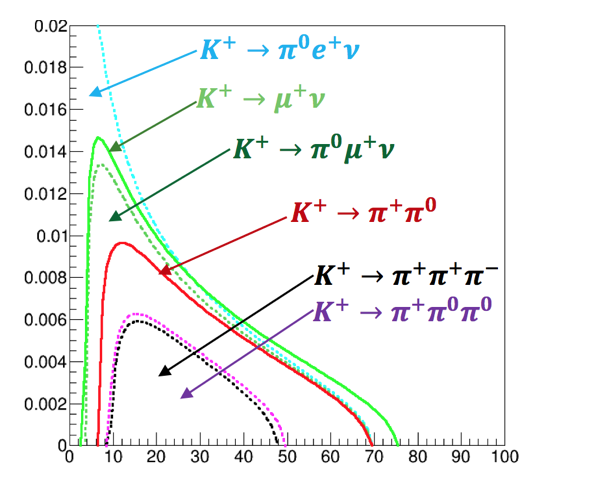}}
\end{minipage}
\caption[]{Distribution of the angle between the decay particle and the kaon beam function of track momentum for data(left) and as expected (right) }
\label{angle}
\end{figure}
Events with only one charged particle are reconstructed in the STRAW spectrometer. Figure \ref{angle} shows the angle between the reconstruction track and the nominal kaon beam direction as a function of momentum.
For comparison, the theoretical curves in the angle - momentum plane expected from kaon decays are also displayed. 
\begin{figure}[h!]
\begin{minipage}{0.50\linewidth}
\centerline{\includegraphics[width=0.7\linewidth]{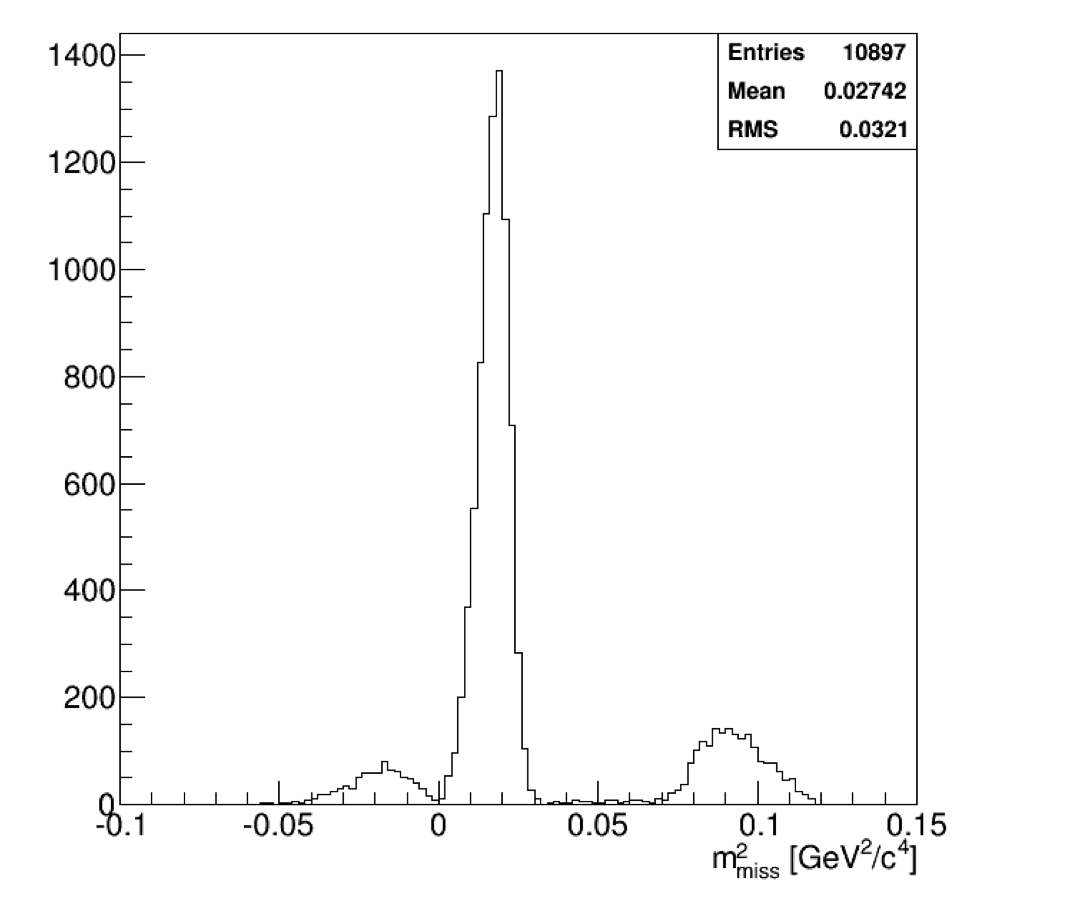}}
\end{minipage}
\hfill
\begin{minipage}{0.50\linewidth}
\centerline{\includegraphics[width=0.7\linewidth]{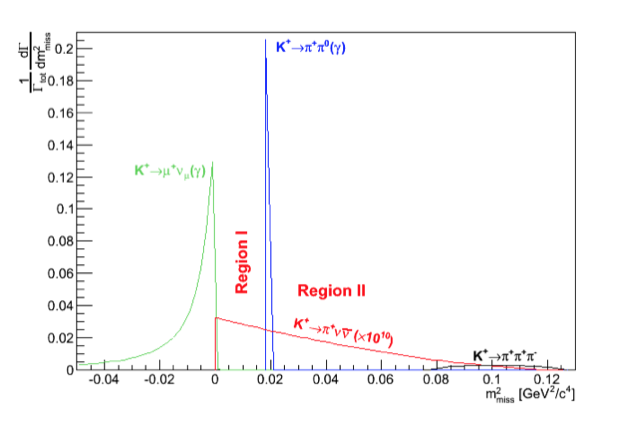}}
\end{minipage}
\caption[]{$m^2_{miss}$ in the hypothesis of kaon to pion decay (left), expected values (right) }
\label{mm2}
\end{figure}
It is possible to see also a 75 GeV/$c$ momentum component, compatible with scattered beam particles reaching the straw spectrometer.
Additional cuts are applied to obtain the $m^2_{miss}$ distribution shown in figure \ref{mm2}.

The most significant requirements are the presence of a kaon track tagged my CEDAR (a five-fold coincidence in the KTAG), the position of the decay vertex in the fiducial region and the momentum of the decay particle between 15 and 35 GeV/$c$. The beam spectrometer GTK is not yet used.

The signal region (section \ref{subsec:exp}) for $K^+ \rightarrow \pi^+\nu \bar{\nu}$ is defined to be $0 < m^2_{miss} < 0.01 ~$GeV$^2/c^4 $ (Region I), and $0.026 < m^2_{miss} < 0.068 ~$GeV$^2/c^4$ (Region II).

Background events enter in  Region I due to resolution tails. A factor of four improvement in the resolution is expected after adding the beam spectrometer information.
Because photon rejection has not been applied yet in this analysis, decays with photons appear in Region II as well as kaon decays with electrons and resolution tails.

\section{Summary}
The NA62 experiment has been taken data for two month in 2014. All sub-detectors have been commissioned and successfully operated with good performance. Data analysis is in progress.

\end{document}